\documentclass[]{jetc2017abstract}
\usepackage{graphicx}
\usepackage{times}

\begin{document}

\vbox{\hbox to \textwidth{\begin{tabular}{@{}c@{\hskip3pc}}
        \hbox to 46pt{\vbox to 3pt{\vss\hsize46pt\vss}}\end{tabular}\hss
        \vbox{\hsize37pc\scriptsize\sf\vskip\baselineskip%
            \hfill \textit {14th Joint European Thermodynamics Conference}\par
            \hfill \textit{Budapest, May 21--25, 2017}\par}
        \hskip1pc}
        }
\vskip15mm

\setcounter{page}{73}

\begin{center}
\textbf{\large\uppercase{GENERIC: Review of successful applications and a challenge for the future}}

\addvspace{12pt}\noindent

\underline{Hans Christian \"Ottinger}$^{1,*}$

\addvspace{10pt}\noindent

$^1$ETH Z\"urich, Department of Materials, Polymer Physics, CH-8093 Z\"urich, Switzerland\\
$^*$hco@mat.ethz.ch\\

\end{center}
%\addvspace{20pt}\noindent
\addvspace{4mm}\noindent

\small

\section*{ABSTRACT}

We present and discuss the two-generator framework of nonequilibrium thermodynamics known as GENERIC (``general equation for the nonequilibrium reversible-irreversible coupling''), which is based on geometric concepts and on statistical mechanics. The usefulness and maturity of the framework are illustrated by reviewing a large number of successful applications. Finally, we offer an important challenge for the future.

This abstract is based on the compact reviews \cite{hco121,hco206} presenting modern nonequilibrium thermodynamics to applied scientists and engineers. In the spirit of a continuation of these reviews, we here add some new developments of the last five years.

\subsection*{GENERIC framework}
Time-evolution equations for nonequilibrium systems possess a well-defined thermodynamic structure in which reversible and irreversible contributions are constructed separately. The reversible contribution is assumed to be of the Hamiltonian form (driven by the gradient of energy) and hence requires an underlying geometric structure which reflects the idea that the reversible time evolution should be ``under mechanistic control.'' The remaining irreversible contribution is assumed to be driven by the gradient of a nonequilibrium entropy. Our discussion is based on the GENERIC framework for closed nonequilibrium systems \cite{hco99,hco100,hcobet} (see also the concise summary in \cite{hco167} adapted for our purposes here),
\begin{equation} \label{LMformulation}
  \frac{dx}{dt} = L \cdot \frac{\delta E}{\delta x} +
  M \cdot \frac{\delta S}{\delta x} ,
\end{equation}
where $x$ represents the set of independent variables required for a complete description of a given closed nonequilibrium system, $E$ and $S$ are the total energy and entropy expressed in terms of the system variables $x$, and $L$ and $M$ are certain linear operators, or matrices. The so-called Poisson matrix $L$ and the friction matrix $M$ can also depend on $x$ so that the fundamental evolution equation (\ref{LMformulation}) can be highly nonlinear. The two contributions to the time evolution of $x$ generated by the total energy $E$ and the entropy $S$ in (\ref{LMformulation}) are the reversible and irreversible contributions to dynamics, respectively. Because $x$ typically contains position-dependent fields, such as the local mass, momentum and energy densities of hydrodynamics, the state variables are usually labeled by continuous (position) labels in addition to discrete ones. A matrix multiplication, which can alternatively be considered as the application of a linear operator, hence implies not only summations over discrete indices but also integrations over continuous labels, and the gradient $\delta/\delta x$ typically implies functional rather than partial derivatives. Equation (\ref{LMformulation}) is supplemented by the complementary degeneracy requirements
\begin{equation} \label{LSconsistency}
  L \cdot \frac{\delta S}{\delta x}=0
\end{equation}
and
\begin{equation} \label{MEconsistency}
  M \cdot \frac{\delta E}{\delta x}=0 .
\end{equation}
The requirement that the entropy gradient $\delta S/\delta x$ is in the null-space of the Poisson matrix $L$ in (\ref{LSconsistency}) expresses the reversible nature of the first contribution to the dynamics, irrespective of the particular form of the Hamiltonian. The requirement that the energy gradient $\delta E/\delta x$ is in the null-space of the friction matrix $M$ in (\ref{MEconsistency}) expresses the conservation of the total energy in a closed system by the irreversible contribution to the dynamics. Furthermore, it is required that the matrix $L$ is antisymmetric, whereas $M$ is Onsager-Casimir symmetric (see Section 3.2 of \cite{hcobet} for details) and positive-semidefinite. Positive-semidefiniteness implies the second law of nonequilibrium thermodynamics,
\begin{equation} \label{net2ndlaw}
  \frac{dS}{dt} = \frac{\delta S}{\delta x} \cdot
  M \cdot \frac{\delta S}{\delta x} \geq 0 ,
\end{equation}
which follows directly from the chain rule by using (\ref{LSconsistency}) and (\ref{MEconsistency}). Finally, the Poisson bracket that can be associated with the antisymmetric matrix $L$ is assumed to satisfy the Jacobi identity, which expresses the time-structure invariance of the reversible dynamics \cite{hco102} and can be conveniently and rigorously tested by using symbolic mathematical tools \cite{hco138,KroegerHuetter10}.

Note that energy and entropy are the fundamental concepts of nonequilibrium thermodynamics. In cases where a nonequilibrium temperature can be introduced in a meaningful way, at least locally, one may be able to combine the two generators $E$ and $S$ into a single one (a ``nonequilibrium free energy''). The possibilities and limitations of the single-generator approach have been discussed in the textbook \cite{BerisEdwards} and in detailed comparisons to the double-generator approach \cite{Edwards98,hco110,Beris01}.

A very reassuring feature of the GENERIC framework is that there exists a clear link to nonequilibrium statistical mechanics. By eliminating the fast degrees of freedom, the projection-operator formalism \cite{Grabert,Robertson66,Mori65,Mori65a,Zwanzig61} produces equations of the GENERIC form (see Chapter~6 of \cite{hcobet} and \cite{hco101,hco131,hco173}). The fast degrees result in noise and friction felt by the slow variables, where these two effects are found to be intimately related according to the fluctuation-dissipation theorem (see Section~1.6 of \cite{KuboetalII}). As a result of the projection procedure, well-defined statistical expressions for the thermodynamic building blocks $E,S,L,M$ in (\ref{LMformulation})-(\ref{MEconsistency}) arise (see Section~6.1.4 of \cite{hcobet}). Evaluation of these expressions should be the Holy Grail of computer simulations for nonequilibrium systems \cite{hco173}. The counterpart in equilibrium statistical thermodynamics is the determination of partition functions (or their partial derivatives) by Monte Carlo simulations to obtain thermodynamic information in terms of the free energy.

A cornerstone of nonequilibrium statistical mechanics is the nonequilibrium ensemble. It is a probability density on the larger space of more microscopic states, parametrized by the more macroscopic state variables taking values from a smaller space (to avoid awkward formulations, we simply refer to microscopic and macroscopic states from now on). The ideas of microcanonical, canonical, and mixed ensembles are taken over from equilibrium statistical mechanics, but now with a much larger and less universal set of thermodynamic nonequilibrium variables. The energy $E$ and the Poisson matrix $L$ of the coarse-grained description can be obtained by simply averaging their miscroscopic counterparts by means of the nonequilibrium ensemble. As at equilibrium, the evaluation of the entropy $S$ is a matter of counting microscopic states or, more generally, of properly normalizing the ensemble. The friction matrix $M$ is the only building block that requires dynamic material information. According to the fluctuation-dissipation theorem, it can be obtained from the time-correlation functions of fluctuations. More precisely, one needs to evaluate the time-integral of two-time correlations of the fluctuations of the macroscopic variables resulting from the elimination of fast microscopic degrees of freedom. The explicit expression for the friction matrix is known as the Green-Kubo formula (see, for example, (3.47) or (6.73) of \cite{hcobet} or (3) of \cite{hco173}).

The respective role and potential of Monte Carlo, molecular dynamics, and Brownian dynamics simulations in thermodynamically guided simulations for nonequilibrium systems has been elaborated in \cite{hco173}. Dynamic simulations should run only over a fraction of the characteristic slow time scales, just sufficiently long to evaluate the decay of two-time correlations on the fast time scales. Initial conditions should be obtained by Monte Carlo sampling from nonequilibrium ensembles. Thermostats and similar devices are unnecessary for such short simulations.

\subsection*{Review of applications}
In addition to the many famous applications of linear irreversible thermodynamics (such as diffusion, osmotic pressure, heat conduction, propagation of sound, electrokinetic effects, thermoelectric effects, thermokinetic effects, dielectric relaxation, polarizable media in electromagnetic fields, magneto-plasmas, superfluids, viscoelastic fluids, and chemical reactions), a number of applications of nonlinear irreversible thermodynamics have been compiled in Appendix~E of the textbook \cite{hcobet}. Those advanced applications are predominantly from the field of complex fluids: reptation model for entangled linear polymers \cite{hco117,hco119,hco123,hco129,EslamiGrmela08,hco210,Stephanouetal16}, pompon model for branched polymers \cite{hco130,vanMeerveld02}, polymer blends \cite{hco108,GrmelaBousPal01,DresslerEdwards04,GuGrmela08,GuGrmelaBous08,hco193}, colloidal suspensions \cite{WagnerN01,GrmelaAitKadiLafl98,Keshtkaretal10,Grmelaetal14}, two-phase systems \cite{Hutter01,Hutter02,Espanol01,EspanolThieu03}), relativistic hydrodynamics \cite{hco109,hco111,hco112,hco114,hco180,Duong15}, discrete formulations of hydrodynamics for simulations \cite{hco122,SerranoEspanol01,EspanolRev03,ElleroEspanolFlek03}, and thermodynamically guided simulations \cite{hco142,hco146,hco145,Kroeger04,hco155,hco157,hco182,IlgKroger11,Ilgetal11,hco195} (see also the review article \cite{hco173} offering ``four lessons and a caveat'' for good simulations in the context of nonequilibrium statistical mechanics).

Several basic transport phenomena have been generalized to the nonlinear regime. For example, diffusion through polymeric and nanocomposite membranes has been modeled by means of the GENERIC framework \cite{LiuDeKee05a,LiuDeKee05b}. Also a comprehensive discussion of the multiscale thermodynamics and mechanics of heat flow goes beyond linear irreversible thermodynamics \cite{GrmelaLebDub11}. Significant progress in applying nonequilibrium thermodynamics to increasingly more complex fluids has, for example, been made with rheological modeling of suspensions of red blood cells \cite{GuGrmelaBous10}, a biological problem of obvious importance. Thermodynamics has also contributed to the understanding of gas flow in the smallest of channels, as in microfluidics, and of aerodynamics of satellites and space stations in the outer limits of our atmosphere \cite{hco194,StruchtrupTorr10,hco198}. Also conceptual developments in electrochemistry benefit from thermodynamic considerations \cite{hco216}.

Whereas the original development of nonequilibrium thermodynamics has mainly been pushed in the context of complex fluids, the general framework is by no means restricted to fluids. Also crystallization phenomena, including polymer crystallization, have been better understood with the help of the methods of modern nonequilibrium thermodynamics and statistical mechanics \cite{Mukherjeeetal04,Hutter04,HutterRutArm05,Hutter06,vanMeerveldetal08,MukherjeeBeris10,BaigEdwards10a,BaigEdwards10b}. Plasticity and viscoplastic solids are further topics in which important issues have been clarified by means of thermodynamics \cite{hco176,HutterTervoort08b,HutterTervoort08,hco192,Mielke11,HuttervBremen12,HutterSvendsen12,HutterTervoort15,Peshkovetal15}. By combining thermodynamics with a thoughtful characterization of the microstructure, valuable insight into continuum damage mechanics has been gained \cite{HutterTervoort08a}. Structural glasses are another challenging problem in physics and materials science for which nonequilibrium phenomena are widely believed to play an important role. Promising new ways to approach this long-standing challenge are suggested or supported by the GENERIC framework \cite{hco167,hco177,hco196,hco208,hco213,FuerederIlg13,SemkivHutter16}.

Most of the applications of nonequilibrium thermodynamics deal with the modeling of bulk systems. To solve the resulting bulk equations one typically needs boundary conditions. The usefulness of linear irreversible thermodynamics for obtaining boundary conditions has been shown by Waldmann in his pioneering 1967 article \cite{Waldmann67}. Brenner and Ganesan \cite{BrennerGane00} asked the very deep question ``Are conditions at a boundary `boundary conditions'?'' Nonequilibrium thermodynamics actually provides the powerful language for expressing the physics at the boundary consistently \cite{hco162,hco171}, thus going well beyond the mathematics of boundary conditions. An illustrative example is provided by the thermodynamic formulation of wall slip \cite{hco172}. Within linear irreversible thermodynamics, a general description of the dynamics of interfaces has been developed by Bedeaux and coworkers \cite{BedeauxAlbMazur76,Bedeaux86,AlbanoBedeaux87}. The main challenge is to generalize the concept of local equilibrium, which is known to be a key ingredient to the nonequilibrium thermodynamics of bulk systems, to lower-dimensional interfaces \cite{JohannessenBed03,JohannessenGroBed08,GlavatskiyBed09,hco204,hco219}. The analysis of the fully nonlinear thermodynamic behavior of complex interfaces within modern nonequilibrium thermodynamics is a very active field of research \cite{hco188,Sagis10,Sagis10a,Sagis11,hco207,Sagis13,Sagis14,SchweizerSagis14,hco223}.

The theory of quantum dissipation \cite{BreuerPetru,Weiss} will play an increasing role in various branches of quantum technology, such as quantum computing, quantum information processing, quantum communication, quantum cryptography, quantum simulation, quantum metrology, quantum sensing, and quantum imaging. After gaining a deep understanding of its geometric structure, modern nonequilibrium thermodynamics suggests the proper form of the equations also for coupling quantum systems to classical environments \cite{Grabert82,hco199,hco221,hco201}. Some applications of the thermodynamic approach to quantum dissipation are discussed in \cite{hco200,hco209,hco217,hco222}. A most remarkable result is that thermodynamics suggests nonlinearity of quantum master equations, contrary to the almost universal presupposition of linear quantum master equations in present-day applications. This nonlinearity can, for example, be handled by stochastic simulation techniques \cite{hco202,hco205}.

\subsection*{A challenge for the future}
According to GENERIC, time-evolution equations for nonequilibrium systems have a well-defined mathematical structure in which reversible and irreversible contributions are identified separately. An important challenge for the future is how the structure of thermodynamically admissible evolution equations can be preserved under time-discretization, which is a key to successful numerical calculations (space-discretization is straightforward because GENERIC can handle discrete systems, as illustrated by the above-mentioned discrete formulations of hydrodynamics for the purpose of simulations). Why should we bother to establish thermodynamically consistent models and boundary conditions if we cannot preserve their features in the numerical solution techniques, which we almost unavoidably have to resort to in most relevant applications? An analogous challenge is well-known to be met by the symplectic integrators used for Hamiltonian systems \cite{SanzSernaCalvo}. Such symplectic integrators are known for better performance in numerical calculations for Hamiltonian systems and hence provide the motivation for similarly introducing GENERIC integrators to perform better calculations for nonequilibrium systems. GENERIC integrators should preserve the time-invariance of the Poisson bracket expressed by the Jacobi identity, the mutual degeneracy requirements, as well as the symmetry and positive-semidefiniteness of the friction matrix. The important advantages of GENERIC integrators hence are:

\noindent (i) The rigorous conservation of an approximate energy leads to limited fluctuations of the physical energy, even over long periods of time.

\noindent (ii) The strict separation of reversible and irreversible contributions to dynamics.

\noindent (iii) A reliable reproduction of the physical entropy production without any spurious dissipation.

Inspired by Moser's work for Hamiltonian systems (see Section~3(e) of \cite{Moser68}), we develop the requirements to be imposed on GENERIC integrators. We then construct GENERIC integrators for systems with a single dissipative process by Taylor expansion. After discussing some details for two specific examples (harmonic oscillator with friction, two gas containers exchanging volume and energy), we summarize the general features of GENERIC integrators and identify some possible directions for future work.

%\bibliography{P:/latex_personal/bibtex/hcopubs}

\end{document}